\documentclass[12pt]{iopart}

\usepackage{booktabs}
\usepackage{graphicx}
\usepackage{subfigure}
\usepackage{indentfirst}
\usepackage{bm}
\usepackage{multirow}

\begin{document}

\title[]{The effect of single-particle space-momentum angle distribution on two-pion HBT correlation in relativistic heavy-ion collisions by using a multiphase transport model}

\author{Hang Yang, Qichun Feng, Yanyv Ren, Jingbo Zhang*, Lei Huo}

\address{School of Physics, Harbin Institute of Technology, Harbin 150001, China}

\ead{jinux@hit.edu.cn}
\vspace{10pt}

\begin{abstract}
With the string melting version of a multiphase transport(AMPT) model, we analyze the transverse momentum dependence of HBT radius $R_{\rm s}$ and the single-pion angle distribution on the transverse plane, in central Au+Au collisions at $\sqrt{S_{NN}}=19.6, 27, 39, 62.4, 200$ GeV. And base on a series of functions, a numerical connection between these two phenomena has been built. We can estimate the single-pion angle distribution on the transverse plane from the HBT analysis. 
\end{abstract}
\vspace{2pc}
\maketitle
%%%%%%%%%%%%%%%%%%%%%%%%%%%%%%%%%%%%%%%%%%%%%%%%%%%%%%%%%%%%%%%%%%%%%%%%%%%%%%%%%%%%%%%%%%%%%%
\section{Introduction}
The two-pion intensity interferometry also called Hanbury-Brown Twiss(HBT) method, was first performed by Hanbury Brown and Twiss to measure the angular diameter of stars in the 1950s\cite{HanburyBrown:1956bqd}. Then G. Goldhaber, S. Goldhaber, W. Lee and A. Pais extended this method in particle physics to study the angular distribution of identical pion pairs in $\overline{p}+p$ collisions\cite{PhysRev.120.300}. Since the two-pion interferometry has been widely used in high-energy heavy-ion collisions. 

In the Relativistic Heavy Ion Collider, a new state of matter quark-gluon-plasma(QGP) has been found\cite{BACK200528}. It is strongly interacting partonic matter formed by deconfined quarks and gluons under extreme temperature and energy density. This state is similar to the early time of the universe after the big bang\cite{Satz:2000}. And there is a critical end point(CEP) at the boundaries between the QGP and hadronic gas on the QCD phase diagram\cite{PhysRevLett.81.4816}. The RHIC collaboration is searching for CEP by the beam energy scan(BES) program\cite{Aggarwal:2010cw}. The HBT method is a useful tool in these researches, it can give the space-time and dynamical information about the freeze-out state of the source\cite{doi:10.1146/annurev.nucl.49.1.529,WIEDEMANN1999145}.

\begin{figure}[htb]
	\centering
	\includegraphics[scale=0.13]{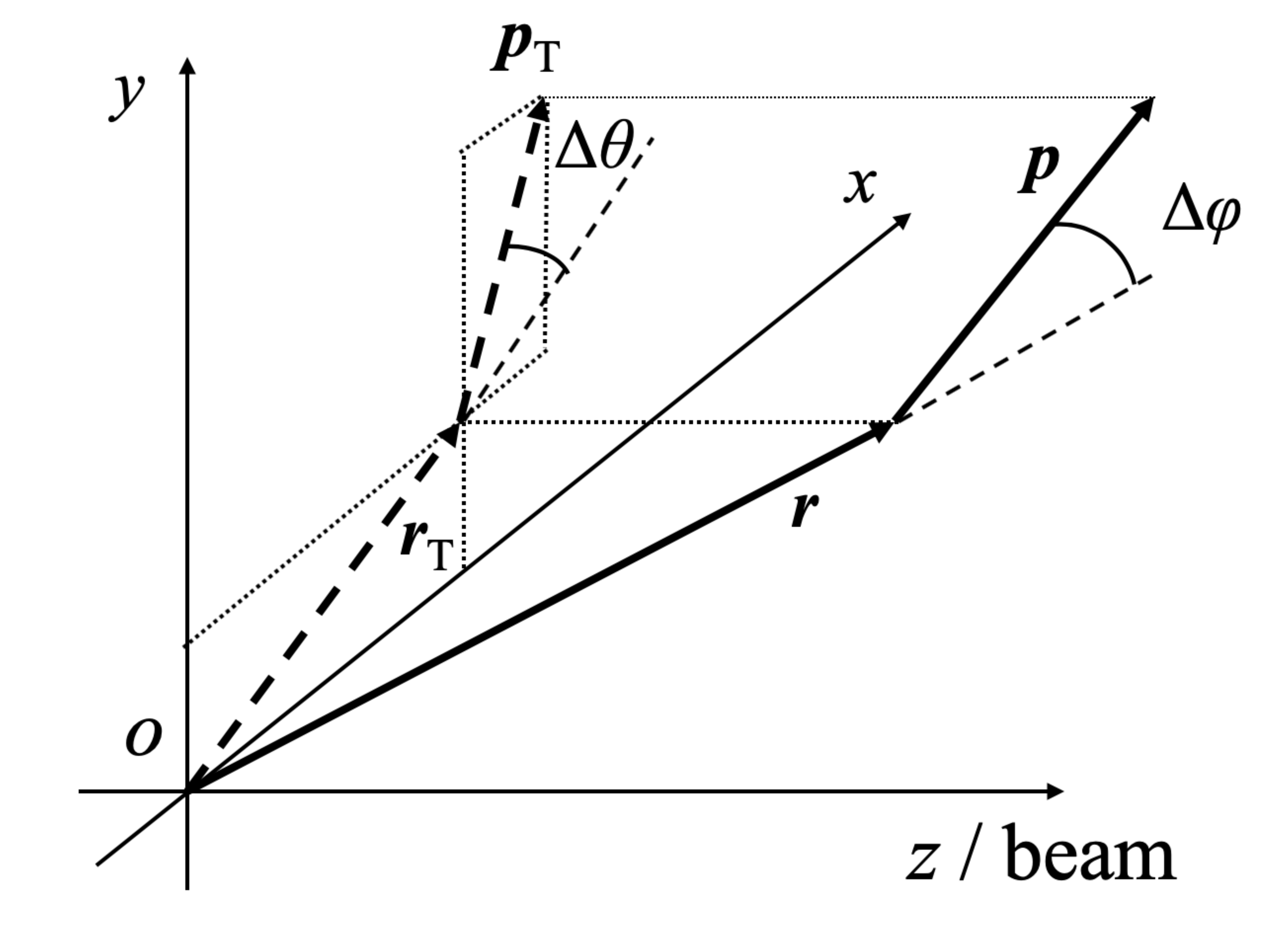}
	\caption{The diagram of the $\Delta\varphi$ and $\Delta \theta$. $\Delta\varphi$ is the angle between $\bm r$ and $\bm p$, and $\Delta \theta$ is the angle between $\bm r_{\rm T}$ and $\bm p_{\rm T}$, at the freeze-out time. The origin is the center of the source.}
	\label{fig_pr}
\end{figure}

Many collaborations have already present some HBT results in different collisions\cite{PhysRevC.71.044906,Kniege_2004,PhysRevD.84.112004,PhysRevC.92.014904}. When analyzed the HBT radii in the same collision energy, the HBT radii show decreases with increasing the transverse momentum of the pair pions. This phenomenon is called transverse momentum dependence of HBT radii, which is attributed to the space-momentum correlation\cite{doi:10.1146/annurev.nucl.55.090704.151533,PhysRevC.53.918}, and the space-momentum correlation is caused by the collective flow\cite{PhysRevLett.53.1219}. We can get more information about the collective flow by studying the transverse momentum dependence of HBT radii, so it is important to study this space-momentum correlation. Hence, we introduce the single-particle space-momentum angle distribution to describe this space-momentum correlation. When the particles freeze out, there will be a finite angle between the radius vector and the momentum vector. Figure~\ref{fig_pr} is the diagram of this angle and its projection angle $\Delta \theta$ on the transverse plane. Then the single-particle space-momentum angle distribution is the quantification of the space-momentum correlation\cite{Yang_2020}. Because this paper is focus on the transverse plane, we only use the angle $\Delta \theta$.

In our study, a multiphase transport (AMPT) model is used to produce particles and calculate the HBT radius. The AMPT model is a hybrid model, which describes the relativistic heavy ion collisions\cite{PhysRevC.72.064901}, it has already been widely used in HBT analysis\cite{PhysRevC.92.014909,PhysRevC.96.044914,Shan_2009}. Then we will focus on the single-particle space-momentum angle $\Delta \theta$ distribution changing with the transverse dependence on the transverse plane. The HBT radius $R_{\rm s}$ is directly related to the transverse size of the pion source\cite{2011328}, so we attempt to build a numerical connection between the $\Delta \theta$ distribution and the transverse dependence of HBT radius $R_{\rm s}$.

This paper is structured as follows. Sec. 2 briefly introduces the AMPT model and the method used to calculate the HBT radii. In Sec. 3, we calculate the HBT radii for pions in different collision energies. In Sec. 4, a numerical connection has been built between the $\Delta \theta$ angle distribution and the transverse momentum dependence of $R_{\rm s}$. Finally, we summarize our conclusions in Sec. 5.

%%%%%%%%%%%%%%%%%%%%%%%%%%%%%%%%%%%%%%%%%%%%%%%%%%%%%%%%%%%%%%%%%%%%%%%%%%%%%%%%%%%%%%%%%%%%%%
\section{AMPT model and methodology}

The AMPT model is a hybrid model with the initial particle distributions generated by the heavy ion jet interaction generator (HIJING) model. The AMPT model contains two versions, the default AMPT model and the string melting AMPT model. The version we used is the string melting AMPT model, which can give a good description of the two-pion correlation function\cite{PhysRevC.72.064901,PhysRevLett.89.152301}.The string melting AMPT model consists of four main components: the initial conditions, partonic interactions, conversion from the partonic to the hadronic matter, and hadronic interactions.

The Correlation After Burner (CRAB) code is used to calculate the two-pion correlation functions\cite{CRAB:2006}. The code is based on the formula
\begin{equation}
	C(\bm{q},\bm{K})=1+\frac{\int {\rm d}^4x_1 {\rm d}^4x_2 S_1(x_1,{\bm p}_2) S_2(x_2,{\bm p}_2)
		{\left|\psi_{\rm{rel}} \right|}^{2}}
		{\int {\rm d}^{4}x_{1}{\rm d}^{4}x_{2}S_{1}(x_{1},{\bm p}_{2})S_{2}(x_{2},{\bm p}_{2})},
\end{equation}
where ${\bm q}={\bm p}_1-{\bm p}_2$, ${\bm K}=({\bm p}_1+{\bm p}_2)/2$, and $\psi_{\rm{rel}}$ is the two particle wave function. $S(x,{\bm p})$ is the emission function of the single particle, and it describes the probability of emitting a particle with momentum $\bm p$ at space-time point $x$.
In further discussion, we neglect the Coulomb interaction and strong interactions between pions, and we focus on the mid-rapidity range($-0.5<\eta<0.5$).

\begin{figure}[htb]
	\centering
	\includegraphics[scale=0.11]{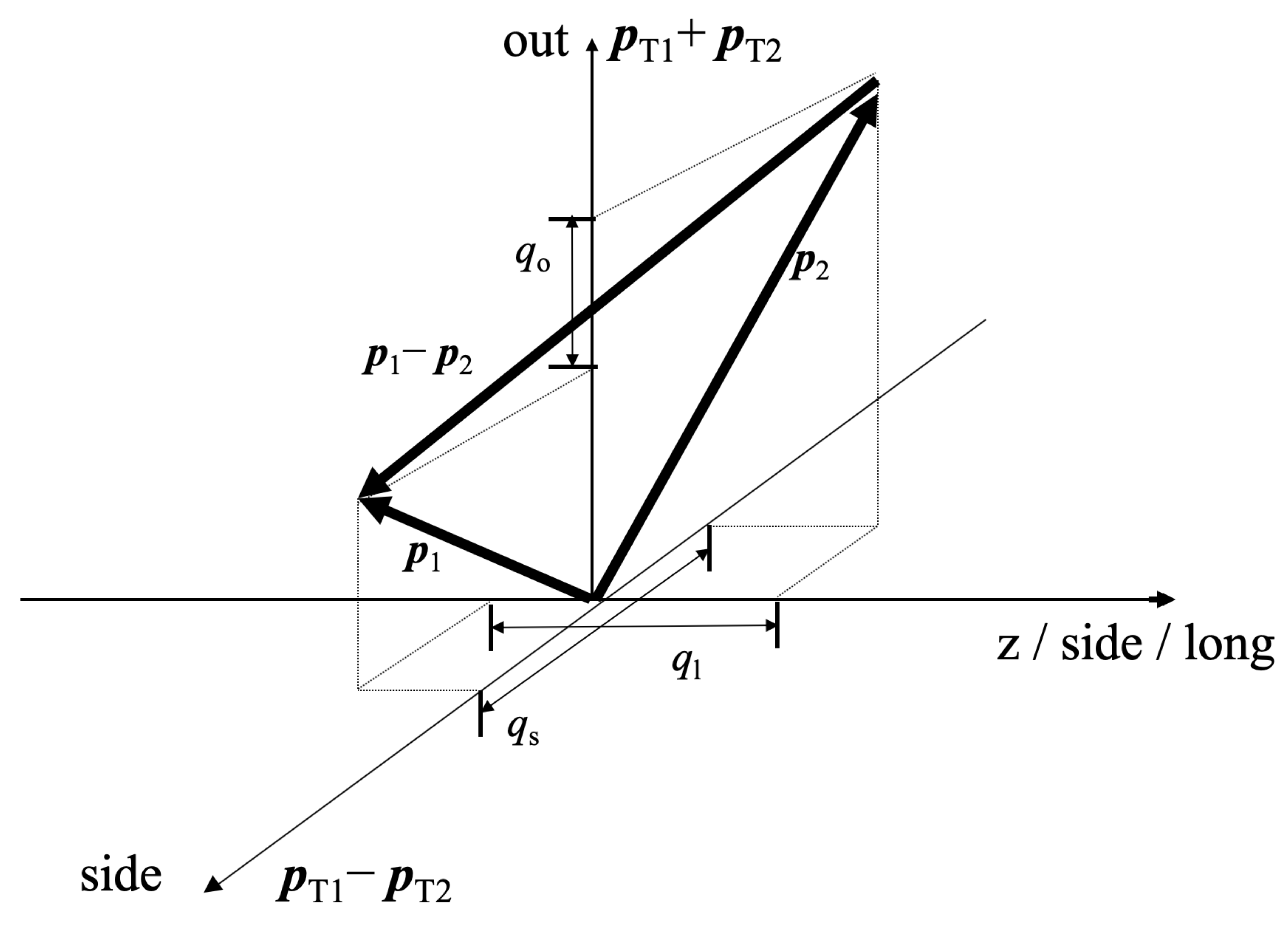}
	\caption{The diagram of `out-side-long'(o-s-l) coordinate system.}
	\label{fig_o-s-l}
\end{figure}

The HBT three-dimensional correlation function can be written as\cite{Alexander_2003}
\begin{equation}
\label{fit_function}
	C(\bm{q},\bm{K})=1+\lambda {\exp}
		{[-q_{\rm o}^2 R_{\rm o}^2(\bm K)-q_{\rm s}^2 R_{\rm s}^2(\bm K)-
		q_{\rm l}^2 R_{\rm l}^2(\bm K)]},
\end{equation}
where $\lambda$ is the coherence parameter. The $R_{\rm o}$, $R_{\rm s}$ and $R_{\rm l}$ are HBT radii in the `out-side-long' coordinate system, the o, s, and l are indicated the directions, shown in Figure~\ref{fig_o-s-l}. The longitudinal direction is along the beam direction, and the transverse plane is perpendicular to the longitudinal direction. In the transverse plane, the momentum direction of pair particles is the outward direction. The direction perpendicular to the outward direction is referred to as the sideward direction. The HBT radii can be calculated by using equation~(\ref{fit_function}) to fit the HBT correlation function generated from the CRAB code.

%%%%%%%%%%%%%%%%%%%%%%%%%%%%%%%%%%%%%%%%%%%%%%%%%%%%%%%%%%%%%%%%%%%%%%%%%%%%%%%%%%%%%%%%%%%%%%%%%%%
\section{Transverse momentum dependence of the HBT radius}

In the HBT analysis, $R_{\rm s}$ is directly related to the transverse size of the emission source,
while $R_{\rm o}$ and $R_{\rm l}$ are influenced by the source lifetime and the velocity of the particles\cite{PhysRevC.53.918}. So $R_{\rm s}$ is more important in our research, so we only focus on the transverse momentum dependence of the $R_{\rm s}$.

\begin{figure}[htb]
	\centering
	\includegraphics[scale=0.1]{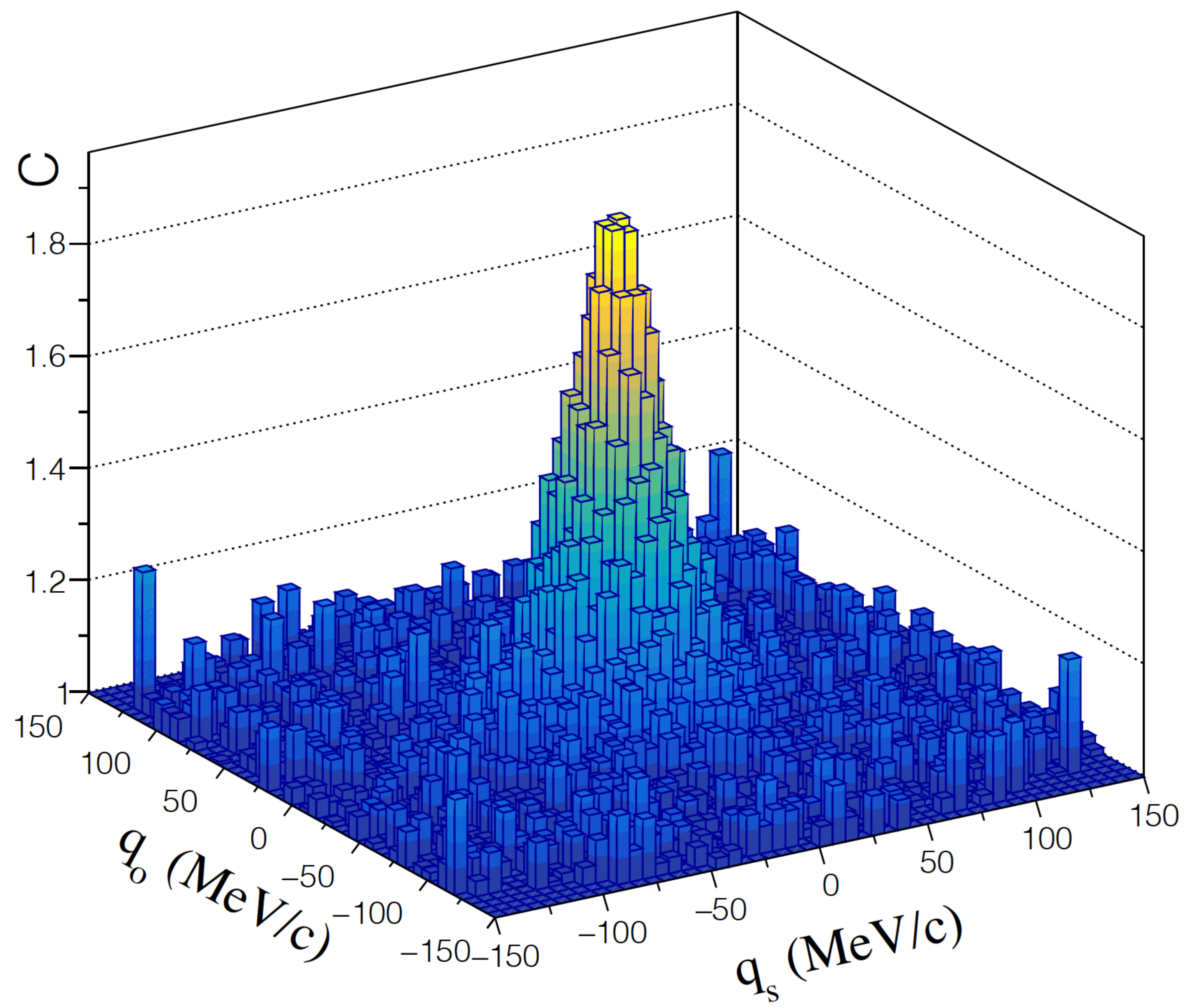}
	\caption{Correlation function in $q_{\rm o}$ and $q_{\rm s}$ directions of the central Au+Au collisions at $\sqrt{S_{NN}}=200$ MeV for the AMPT model. The $K_{\rm T}$ range is 175-225 MeV/c, and the $q_{\rm l}$ range is 0-6 MeV/c.}
	\label{fig_c2}
\end{figure}

We use melting AMPT model generate central Au+Au collision events at $\sqrt{S_{NN}}=19.6, 27, 39, 62.4, 200$ GeV, these energies are chosen from the BES energies, and the impact parameter is 0 fm. The Coulomb interaction and strong interactions are neglected, so the three kinds of pions can be treated as one kind. The correlation function of pions are calculated in different transverse momentum by the Crab code. An example of the correlation function of pions is shown in Figure~\ref{fig_c2}.

\begin{figure}[htb]
	\centering
	\includegraphics[scale=0.3]{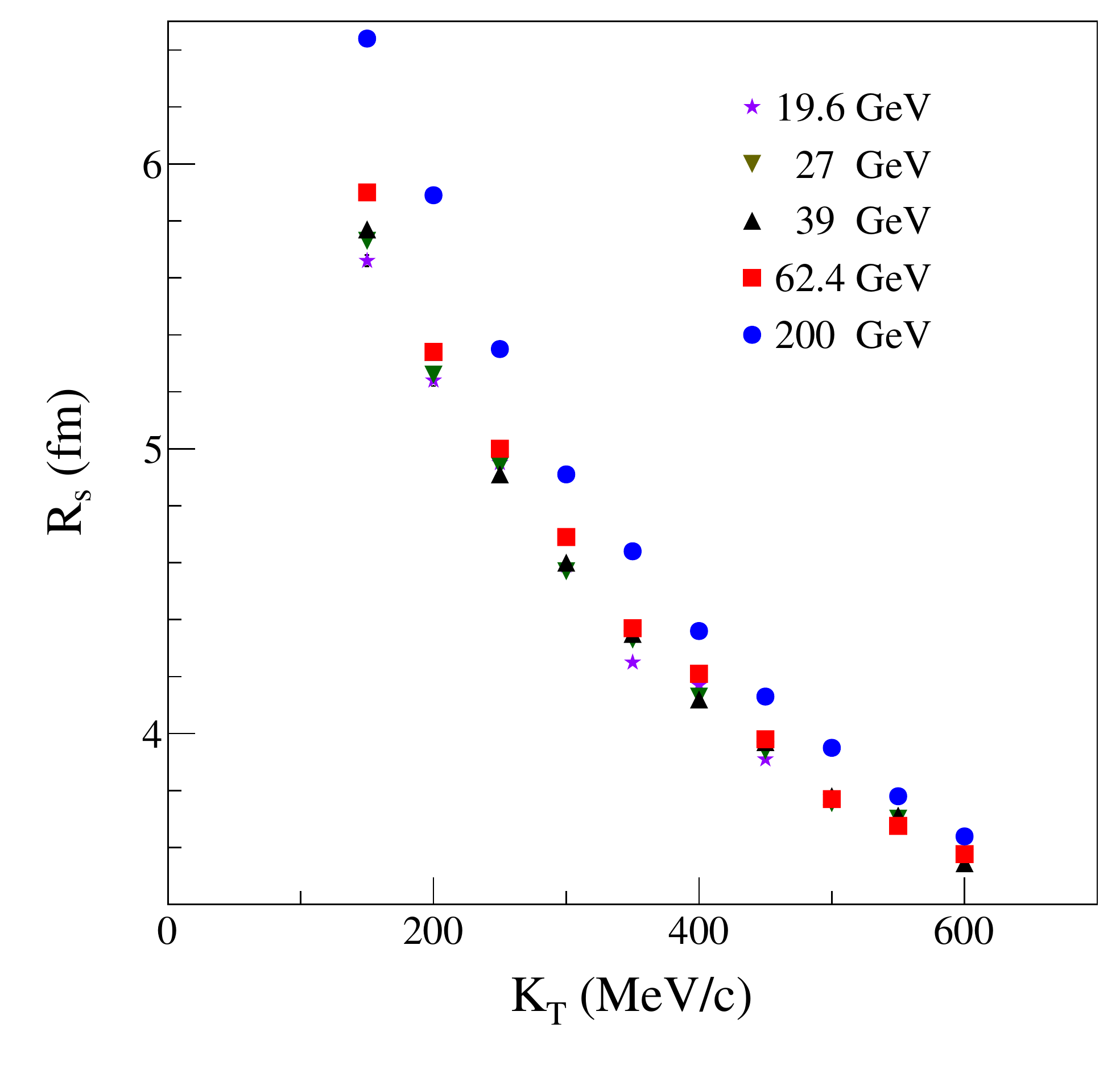}
	\caption{Transverse momentum dependence of $R_{\rm s}$ in AMPT model.}
	\label{fig_kt_r}
\end{figure}

Then we use the equation~(\ref{fit_function}) to fit these HBT correlation functions, and get the values of the $R_{\rm s}$ of different transverse momentum in different collision energies, they are shown in Figure~\ref{fig_kt_r}.

\begin{figure}[htb]
	\centering
	\includegraphics[scale=0.3]{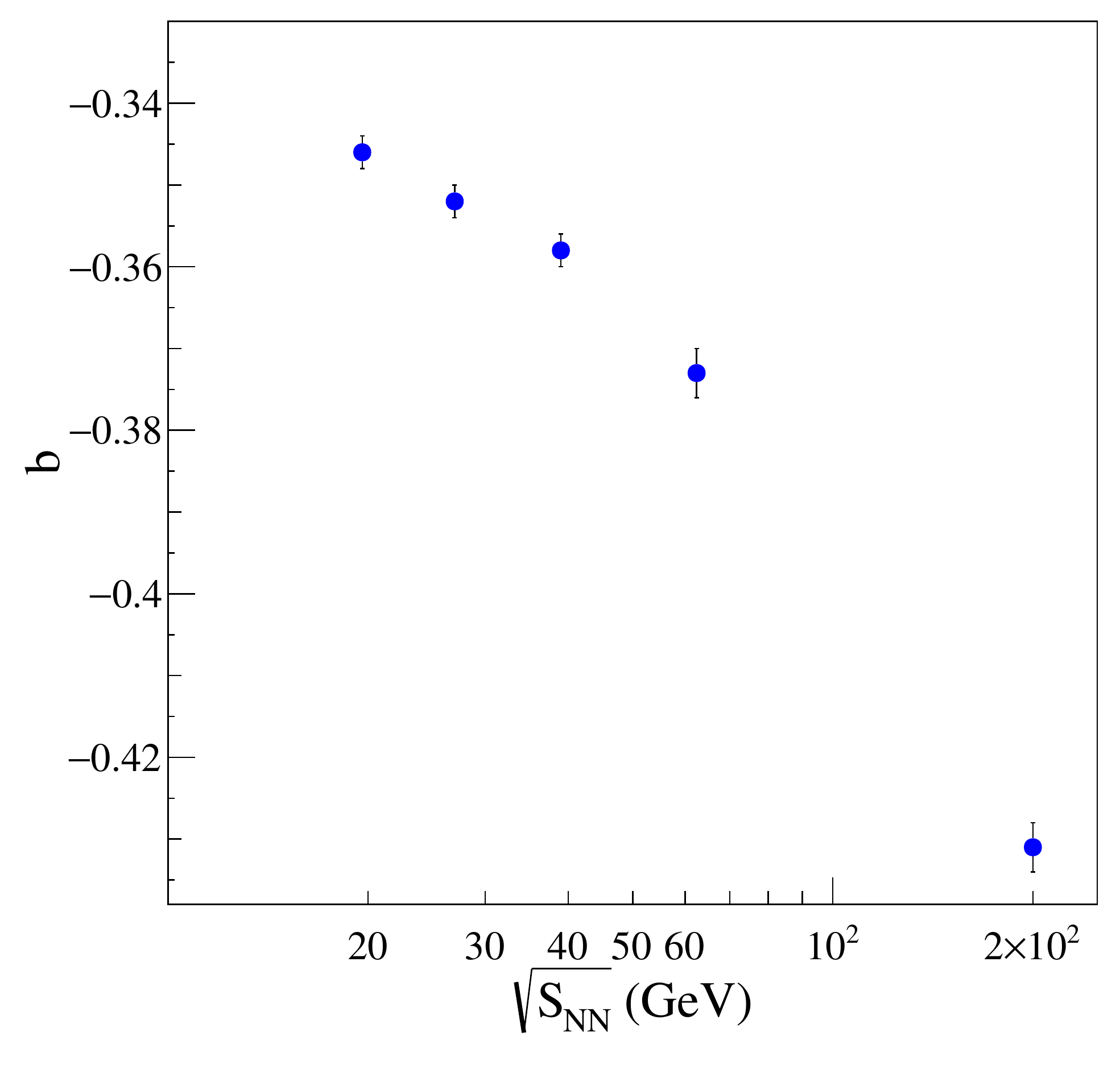}
	\caption{Collision energy dependence of parameter $b$ in AMPT model.}
	\label{fig_snn-b}
\end{figure}

In different collision energies, the $R_{\rm s}$ always shows the decreases with the increasing the transverse momentum of pairs, and they show different strength of $K_{\rm T}$ dependence of $R_{\rm s}$. The values of $R_{\rm s}$ with higher $K_{\rm T}$ are closer too each other. We can fit the HBT radii by
\begin{equation} 
	R = a K_{\rm T}^{b},
\end{equation} 
where parameter $a$ is a common constant, and parameter $b$ reflects how the HBT radii change with the $K_{\rm T}$, it can describe the strength of $K_{\rm T}$ dependence of HBT radii. The collision energy dependence of parameter $b$ is shown in Figure~\ref{fig_snn-b}.

In Figure~\ref{fig_snn-b}, with the increase of collision energy, and the values of parameter $b$ for $R_{\rm s}$ become larger, i.e., a larger $|b|$, indicates a more prominent $K_{\rm T}$ dependence. Thus, we only focus on the dependence of the $cos(\Delta \theta)$ distribution on the parameter $b$.
%%%%%%%%%%%%%%%%%%%%%%%%%%%%%%%%%%%%%%%%%%%%%%%%%%%%%%%%%%%%%%%%%%%%%%%%%%%%%%%%%%%%%%%%%%%%%%%%%%%
\section{Transverse space-momentum angle distribution}

\begin{figure}[!htb]
	\centering
	\subfigure[]
	{
		\includegraphics[width=0.4\textwidth]{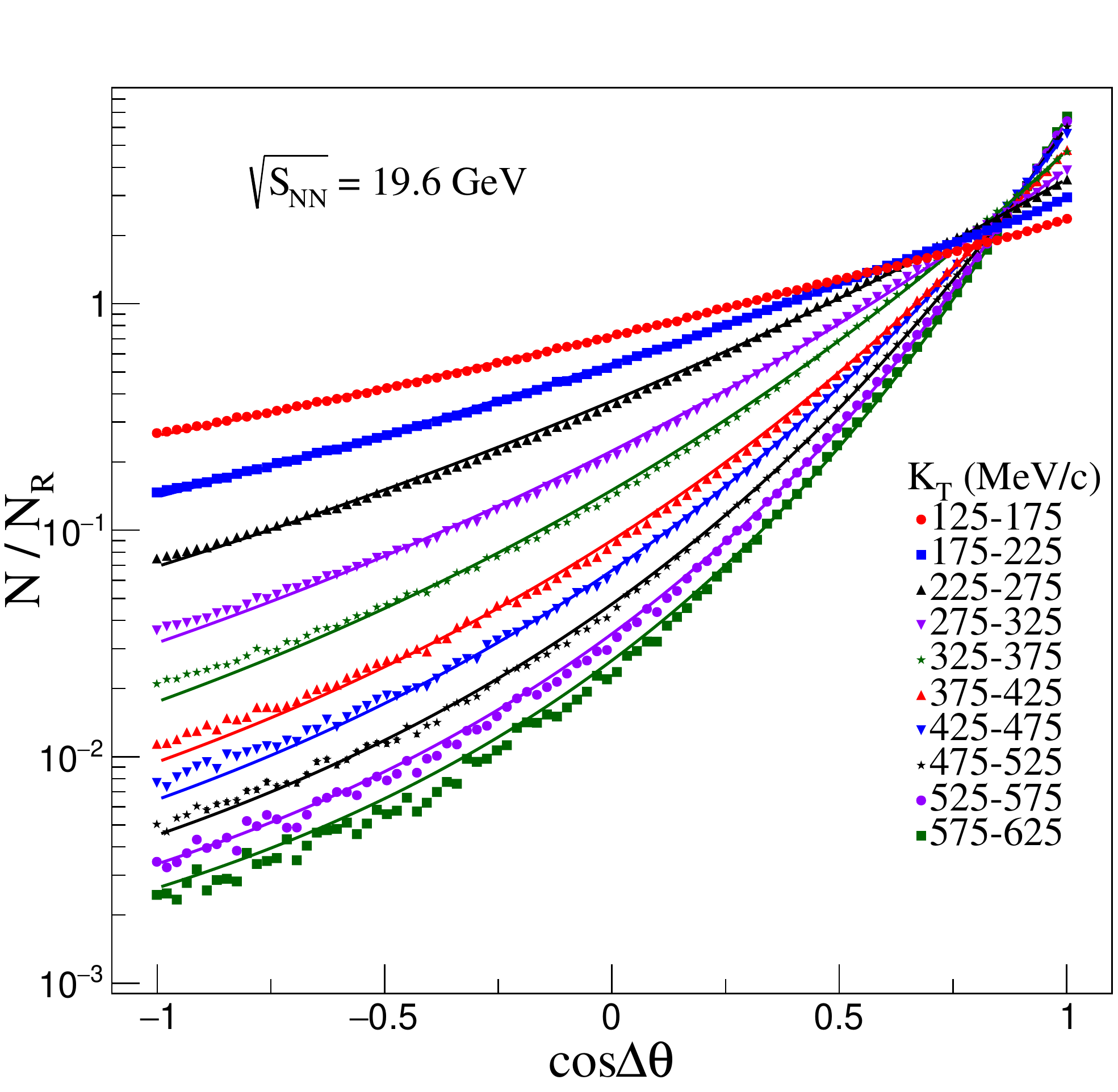}
	}
	\subfigure[]
	{
		\includegraphics[width=0.4\textwidth]{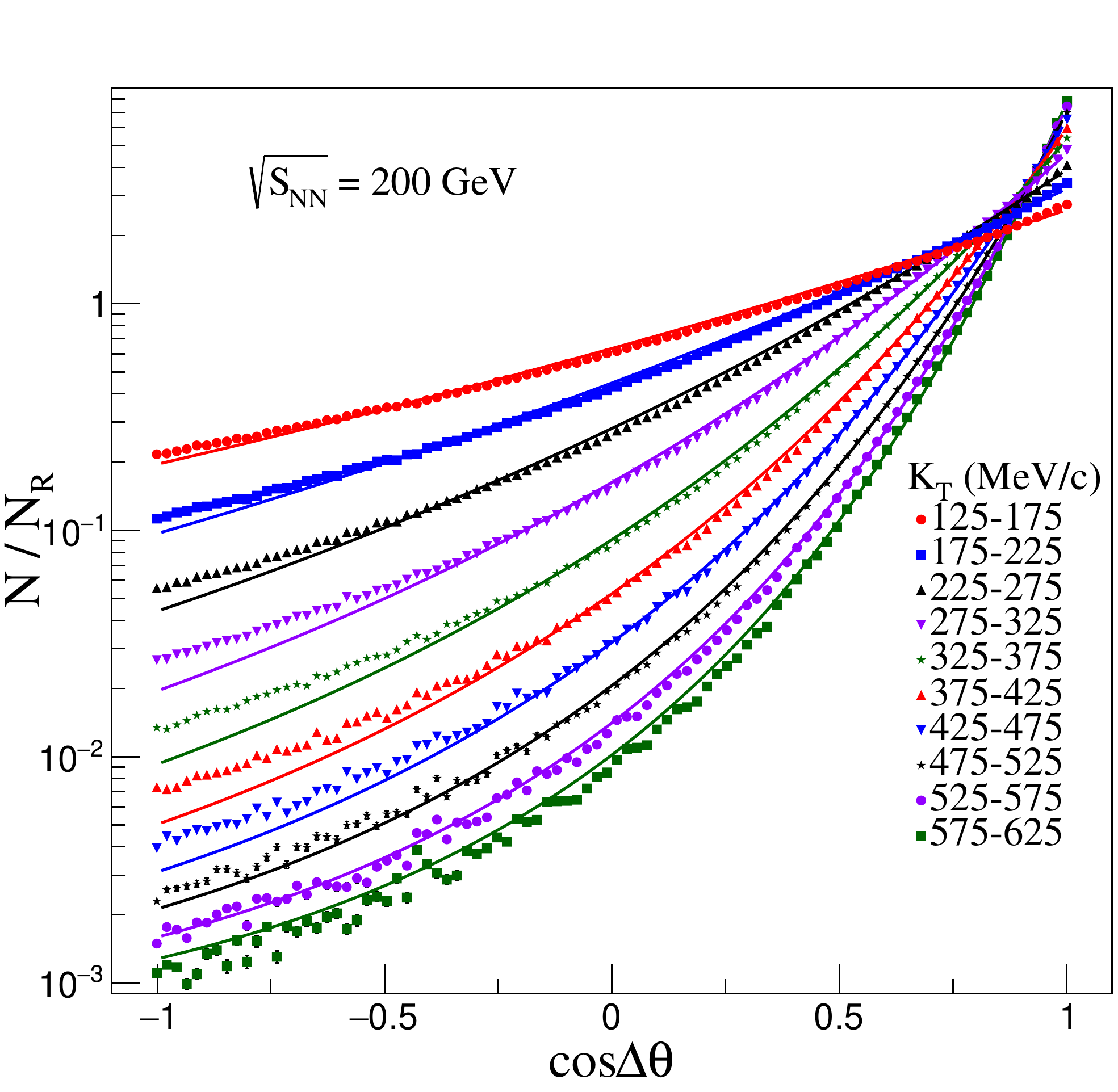}
	}
	\caption{Normalized $\cos(\Delta\theta)$ distribution for melting AMPT model.}
	\label{fig_costheta}
\end{figure}

The single-particle space-momentum angle distribution in transverse plane, also called $\Delta \theta$ distribution, can directly cause the transverse momentum $K_{\rm T }$ dependence of $R_{\rm s}$\cite{Yang_2020}. We use the pions, which are generated from the AMPT model and used to calculate the HBT correlation function, to obtain the $\cos(\Delta \theta)$ distribution. And we generate pions that have random $\bm p_{\rm T}$ and random $\bm r_{\rm T}$, and use them to calculate the random $\cos(\Delta\theta)$ distribution. Then we use the distribution of $\cos(\Delta\theta)$, which is calculated from the AMPT model and divided by the random $\cos(\Delta\theta)$ distribution, to get the normalized $\cos(\Delta\theta)$ distribution, as shown in Figure~\ref{fig_costheta}.

\begin{figure}[!htb]
	\centering
	\subfigure[]
	{
		\includegraphics[width=0.4\textwidth]{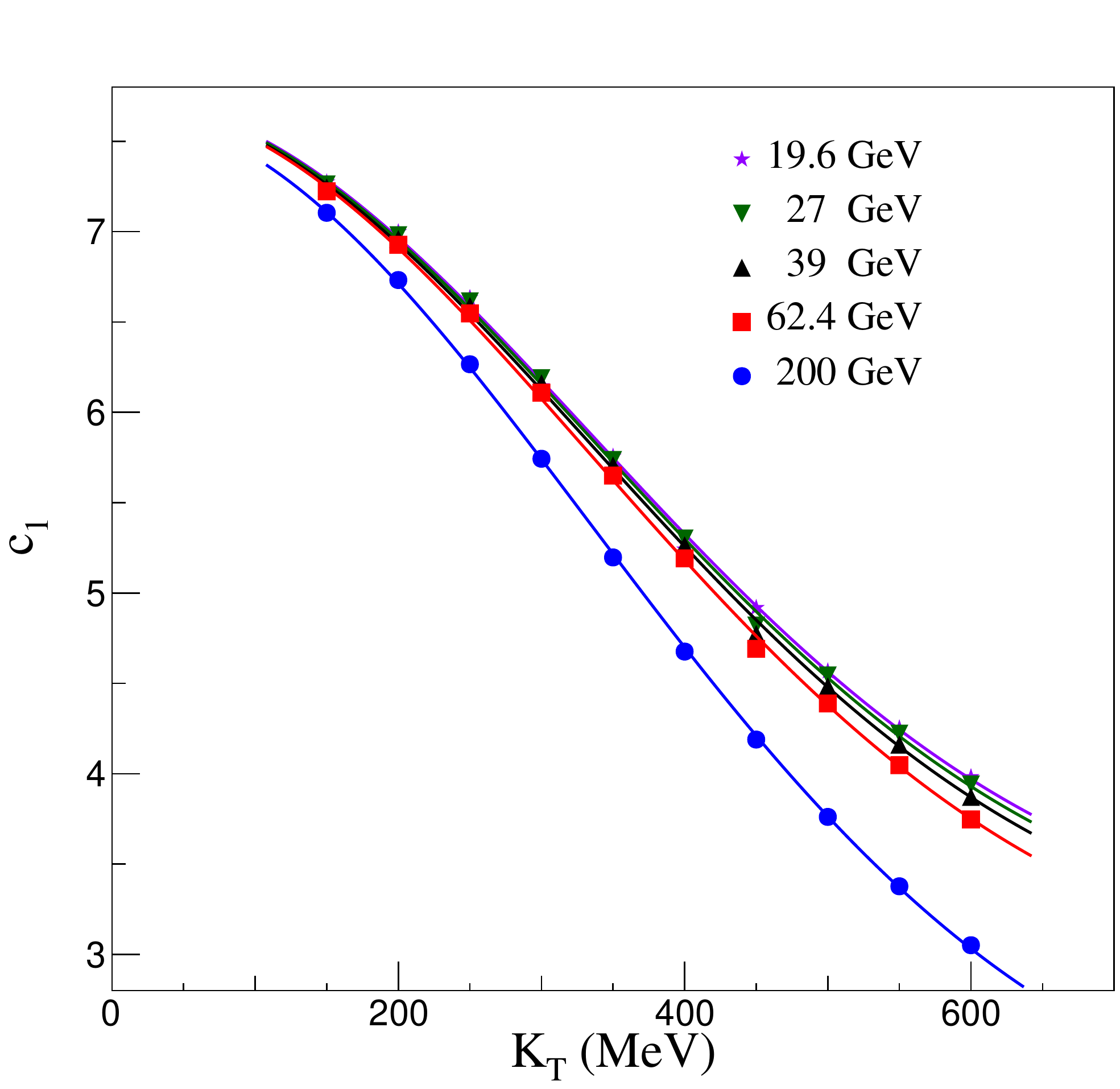}
	}
	\subfigure[]
	{
		\includegraphics[width=0.4\textwidth]{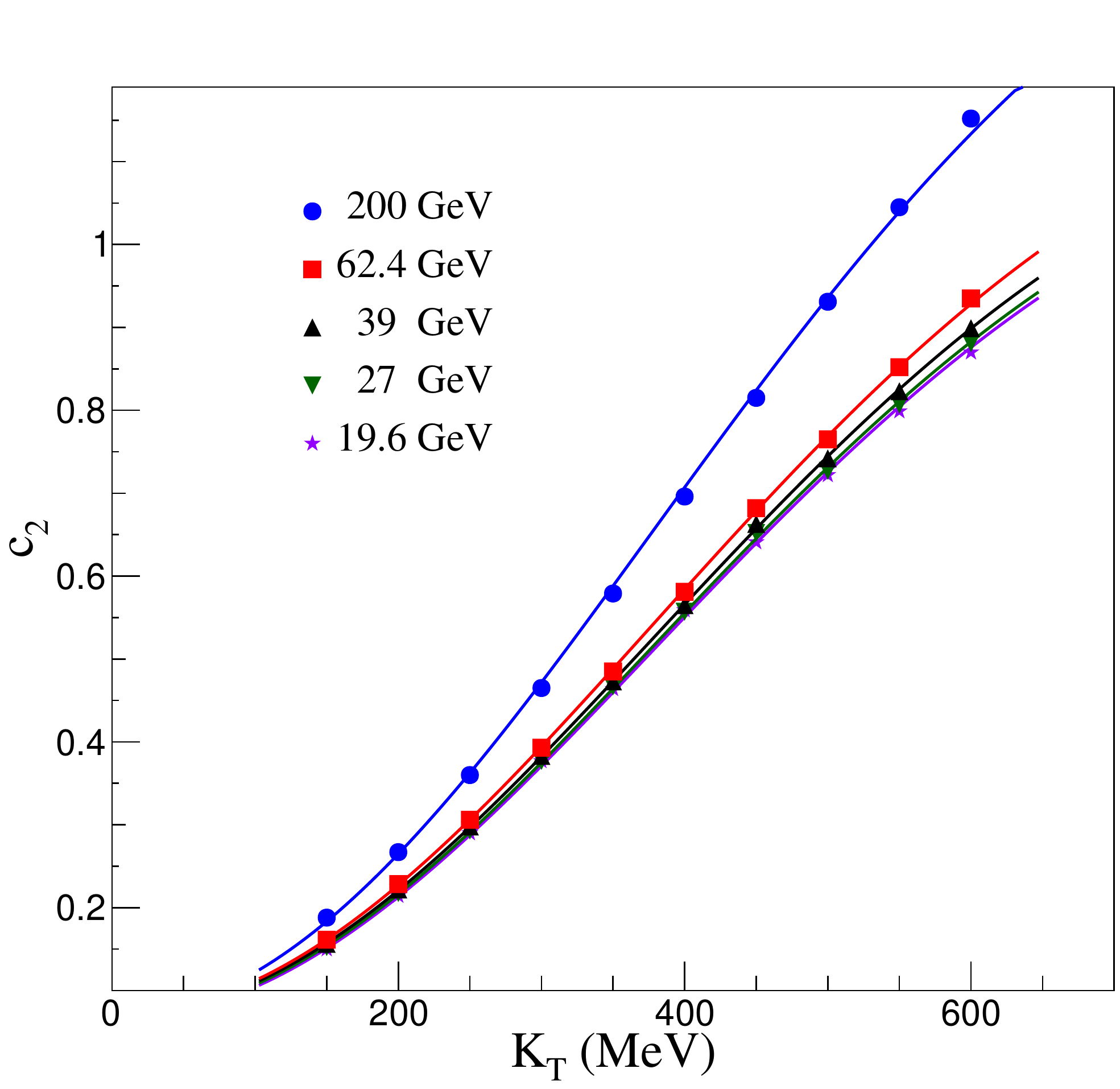}
	}
	\caption{Transverse momentum dependence of fit parameters of $c_{1}$ and $c_{2}$.}
	\label{fit_c1_c2}
\end{figure}

In Figure~\ref{fig_costheta}, with increase the transverse momentum $K_{\rm T}$ and the collision energies, the $\cos(\Delta \theta)$ distribution are closer to the $\cos(\Delta \theta) = 1$. This phenomenon indicates that with the increase of the two pions transverse momentum $K_{\rm T}$ and the collision energies, the momentum direction of single-pion is closer to the space direction. And the transverse momentum dependence of HBT can be explained as the transverse momentum dependence of the normalized single-particle space-momentum angle $\Delta\varphi$ distribution.  The fit functions of the fit lines are
\begin{equation} 
	f=0.0005\exp\bigg\{ c_{1}\exp\Big[c_{2}\cos(\Delta\theta) \Big] \bigg\},
\end{equation} 
where $c_{1}$ and $c_{2}$ are fit parameters, the value 0.0005 is settled by us to get the good fitting results. There are ten bins of $K_{\rm T}$ and five collision energies, so we have ten groups of parameters $c_{1}$ and $c_{2}$ for each energy. Moreover, they can also be fitted by
\begin{eqnarray} 
	c_{1}=k_{1} \exp\Big[-4.5\times(\frac{K_{\rm T}}{1000})^{2} \Big]+j_{1},\\
	c_{2}=k_{2} \exp\Big[-3.5\times(\frac{K_{\rm T}}{1000})^{2} \Big]+j_{2},
\end{eqnarray} 
where $k$ and $j$ are fit parameters, the values -4.5 and -3.5 are also settled by us to get the good fitting results. The fit lines are shown in Figure~\ref{fit_c1_c2}. At the low collision energies, especially $\sqrt{S_{NN}}=19.6, 27, 39$GeV, the collision energies are close to each other, and the strength of transverse flow is also similar. It leads the single-pion space-momentum angle to have a similar distribution in the same transverse momentum section, so the $c_{1}$ and $c_{2}$ values are close to each other. 

\begin{figure}[!htb]
	\centering
	\subfigure[]
	{
		\label{k1_b}
		\includegraphics[width=0.33\textwidth]{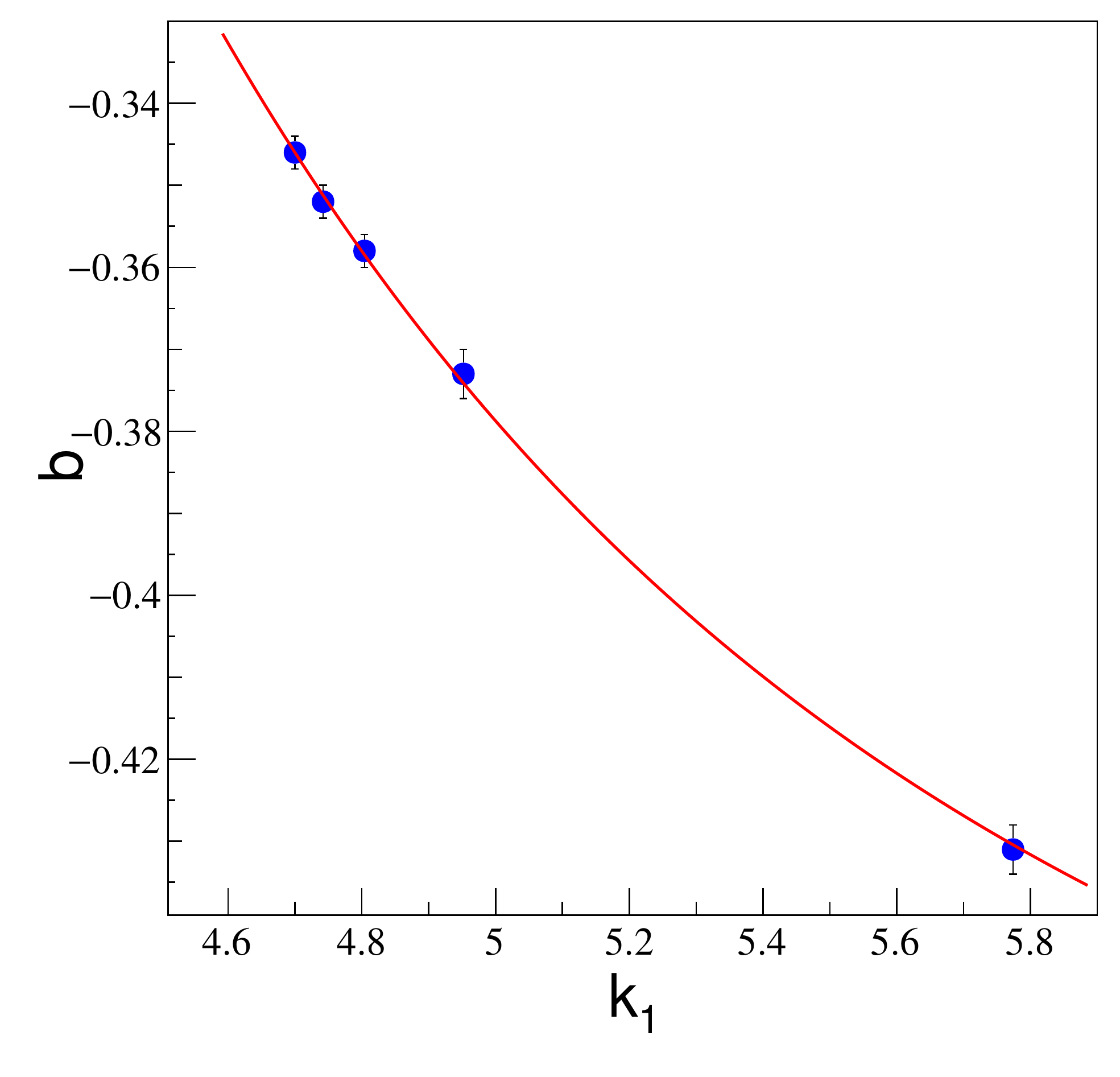}
	}
	\subfigure[]
	{
		\label{j1_b} 
		\includegraphics[width=0.33\textwidth]{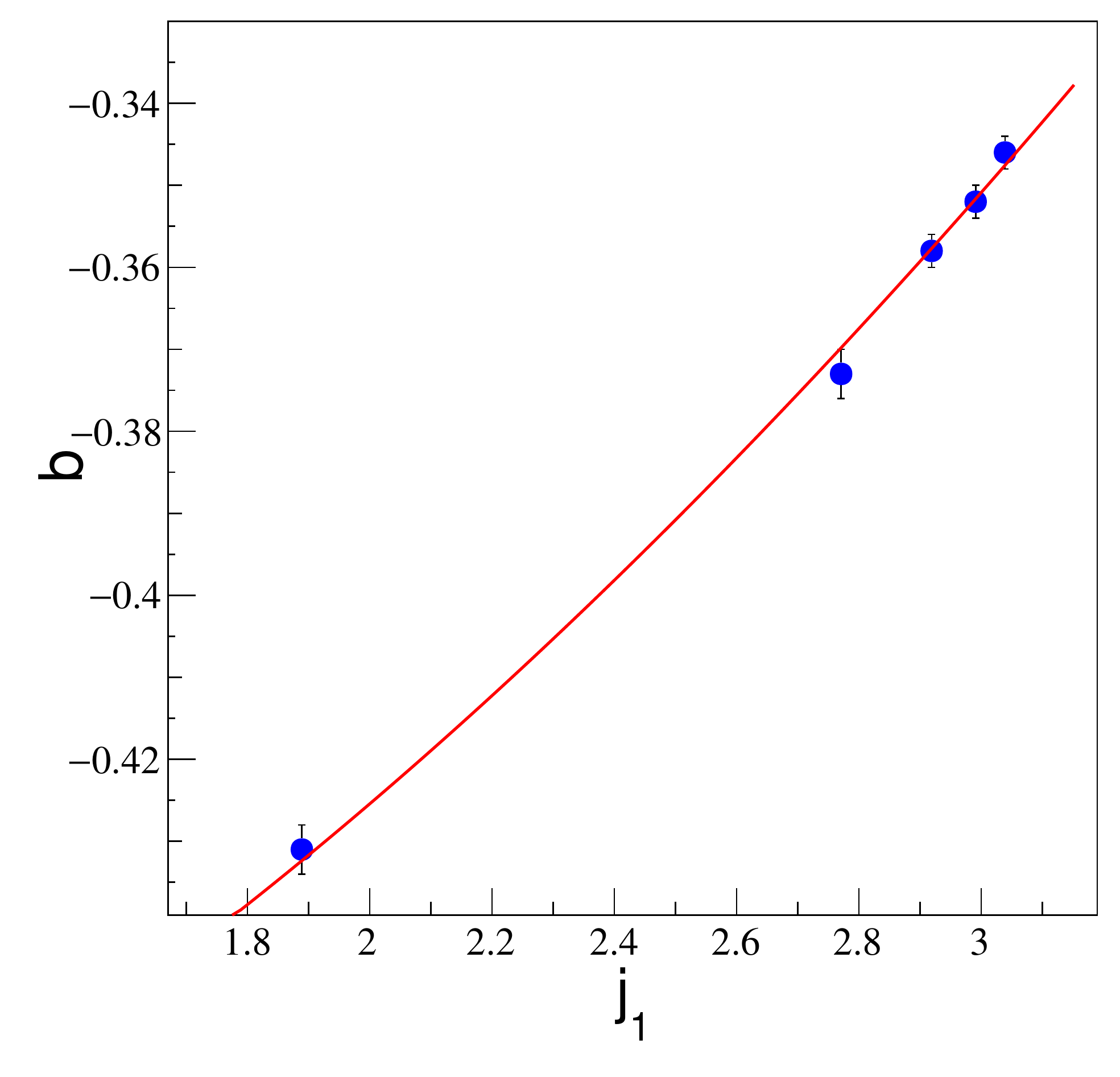}
	}
	
	\subfigure[]
	{
		\label{k2_b}
		\includegraphics[width=0.33\textwidth]{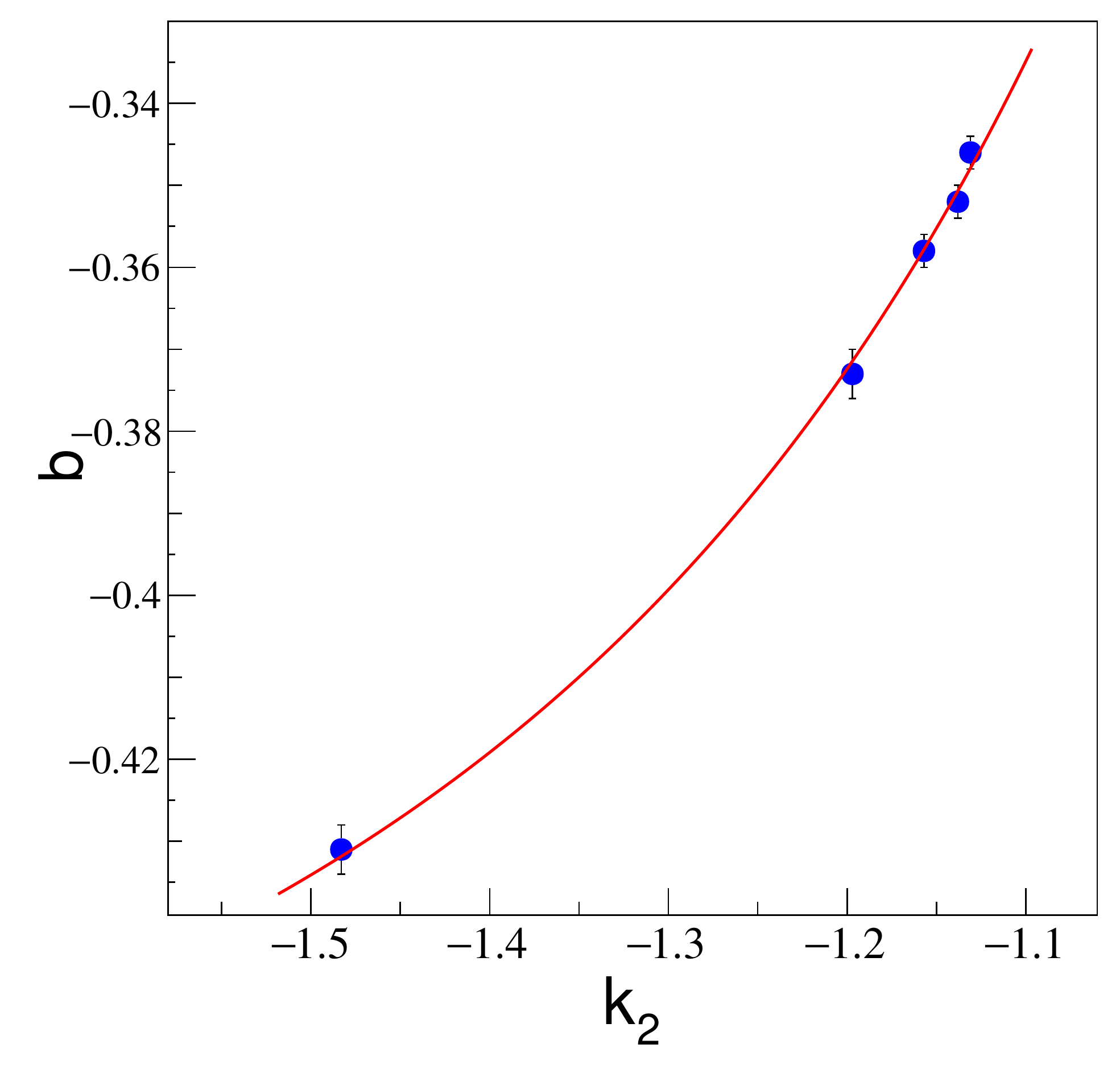}
	}
	\subfigure[]
	{
		\label{j2_b} 
		\includegraphics[width=0.33\textwidth]{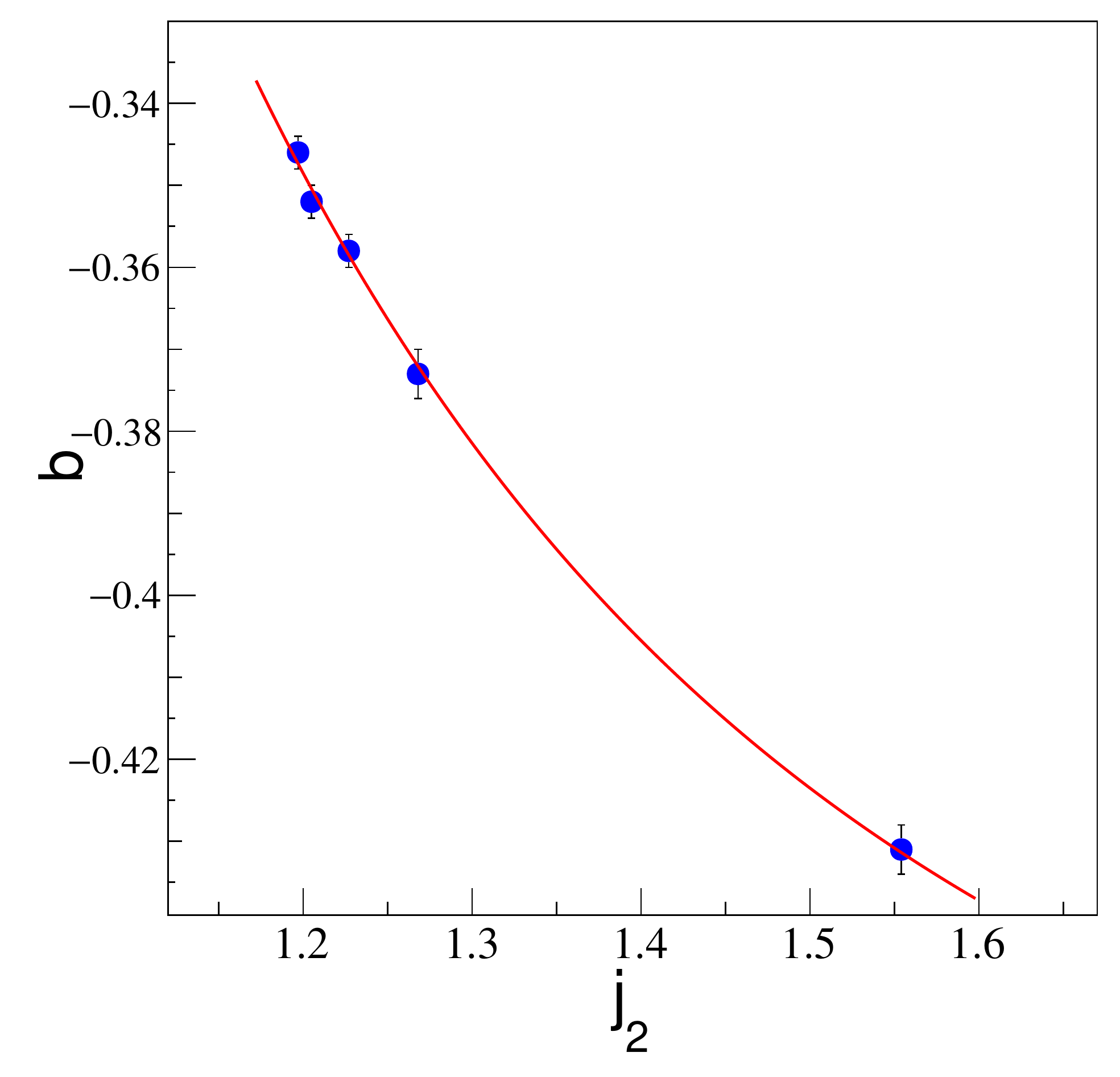}
	}
	\caption{Fit parameters in AMPT model. Parameter $b$ from HBT radii fit function $R = a p_{\rm T}^{b}$, $j_1$, $j_2$ and $k_1$, $k_2$ are from fit function $c_{1}=k_{1} \exp[-4.5\times(\frac{K_{\rm T}}{1000})^{2}]+j_{1}$ and $c_{2}=k_{2} \exp[-3.5\times(\frac{K_{\rm T}}{1000})^{2}]+j_{2}$, $c_1$ and $c_2$ are from normalized space-momentum angle distribution function $f=0.0005\exp\{ c_{1}\exp[c_{2}\cos(\Delta\theta)]\}$, and the red lines are fit lines.}
	\label{fig_b_j_k}
\end{figure}

We plot parameters b, which obtained from the transverse momentum dependence of $R_{\rm s}$, as the functions of parameter of k and j in Figure~\ref{fig_b_j_k}. With the increase of collision energies, the transverse flow will also increase, the pions will be easily frozen out along the space direction, so the $\cos(\Delta \theta)$ distribution will be close to the $\cos(\Delta \theta)=1$. The changing pattern is carried by the fit parameters. The pattern indicated that parameter $b$ has an extremum, so the effect of single-pion space-momentum angle distribution on the HBT radii will be similar at the higher energies, and the transverse dependence of the HBT radii will also be similar at the higher collision energies. We set this minimum value to -0.5 and we can make the numerical connection by the fit results. The red lines depict fits, and the fit functions are

\begin{eqnarray}
	b(k_1) = \mu_{11} |k_1| ^{\mu_{12}}-0.5,\label{bk1}    \\
	b(j_1) = \nu_{11} |j_1| ^{\nu_{12}}-0.5,\label{bj1}    \\
	b(k_2) = \mu_{21} |k_2| ^{\mu_{22}}-0.5,\label{bk2}    \\
	b(j_2) = \nu_{21} |j_2| ^{\nu_{22}}-0.5.\label{bj2}
\end{eqnarray}
where $\mu$ and $\nu$ are fit parameters. The fit parameters values are shown in Table~\ref{muandnu}.  
\begin{table}[!htb]
\begin{indented}
\lineup
\item[]
	\caption[table]{Fit results of $b(k)$ and $b(j)$}\label{muandnu}
	\begin{tabular}{@{}*{4}{c}}
		\br
		{} & $c_1$ &{}&$c_2$\\
		\mr
		$\mu_{11}$ & $60\pm20	$& $\mu_{21}$ & $0.219\pm0.005	$	\\
		$\mu_{12}$ & $-3.9\pm0.2 		$& $\mu_{22}$ & $-3.0\pm0.2  	$ 	\\
		$\nu_{11}$ & $0.023\pm0.002   $& $\nu_{21}$ & $0.265\pm0.009   $ 	\\
		$\nu_{12}$ & $1.7\pm0.1 		$& $\nu_{22}$ & $-3.1\pm0.2  	$	\\
		\br
	\end{tabular}
\end{indented}
\end{table}

With the AMPT model, a numerical connection has been made between the transverse momentum dependence of HBT radius $R_{\rm s}$ and the single-pion space-momentum $\cos(\Delta \theta)$ distribution. When we obtain a series of data of $R_{\rm s}$ in different $K_{\rm T}$ regions, we can estimate the $\cos(\Delta \theta)$ distribution as a function of $K_{\rm T}$. With further research, the values settled by us can also be improved by more accurate fitting results.

%%%%%%%%%%%%%%%%%%%%%%%%%%%%%%%%%%%%%%%%%%%%%%%%%%%%%%%%%%%%%%%%%%%%%%%%%%%%%%%%%%%%%%%%%%%%
\section{Conclusions}
The transverse momentum dependence of HBT radii is caused by the space-momentum correlation, and we use the single-particle space-momentum angle $\cos(\Delta \theta)$ distribution to quantify this correlation. Thus the transverse momentum dependence of HBT radii can be explained by the transverse momentum dependence of the single-particle space-momentum angle $\cos(\Delta \theta)$ distribution. With the string melting AMPT model, we calculate the transverse momentum of HBT radius $R_{\rm s}$ in several collision energies, and the results show that, with the increase of the collision energies, the decrease of the HBT radii with the pair momentum is more prominent. Then we calculate the single-particle space-momentum angle $\cos(\Delta \theta)$ distribution in each pair momentum section and collision energy. The results show that, with the increase of the collision energies and the pair momentum, the freeze-out directions of the pions are closer to the radius directions. We use several fit functions to build a connection between these distributions and the $R_{\rm s}$. With this connection, we can describe how the space-momentum correlation changing with the increasing collision energies, and we can get information about the final stage of the Au+Au collision at freeze-out time. With further research, the HBT analysis may help us learn about the collective flow and the freeze-out stage of the collisions.
%%%%%%%%%%%%%%%%%%%%%%%%%%%%%%%%%%%%%%%%%%%%%%%%%%%%%%%%%%%%%%%%%%%%%%%%%%%%%%%%%%%%%%%%%%%%
\section*{References}
\bibliography{references}

\end{document}